\documentclass[a4paper, UKenglish, anonymous,cleveref, autoref,thm-restate,nolineno]{socg-lipics-v2021}

\PassOptionsToClass{anonymous}{lipics-v2021}

\usepackage[linesnumbered, ruled, vlined]{algorithm2e}
\usepackage{cleveref}
\usepackage{multicol}
\usepackage{lineno}
\usepackage{graphicx} 

\usepackage{tikz}
\usepackage{subcaption}
\usetikzlibrary{trees, positioning}
\usepackage{caption}
\usetikzlibrary{fit}
\usepackage{xcolor}

\crefname{algocf}{algorithm}{algorithms}
\Crefname{algocf}{Algorithm}{Algorithms}

\title{Dynamic Maximal Independent Set using Predictions} 

\titlerunning{Dynamic MIS using predictions} 

\author{Shahbaz Khan}{Department of Computer Science and Engineering, Indian Institute of Technology Roorkee, India\and \url{https://shahbazk.github.io/}}{shahbaz.khan@cs.iitr.ac.in}{[https://orcid.org/0000-0001-9352-0088]}{}

\author{Anshul Kumar}{Department of Computer Science and Engineering, Indian Institute of Technology Roorkee, India}{anshul_k@cs.iitr.ac.in}{}{}

\authorrunning{S. Khan and A. Kumar} 

\Copyright{Shahbaz Khan and Anshul Kumar } 



\category{} 

\relatedversion{} 

\nolinenumbers 

\EventEditors{John Q. Open and Joan R. Access}
\EventNoEds{2}
\EventLongTitle{42nd Conference on Very Important Topics (CVIT 2016)}
\EventShortTitle{CVIT 2016}
\EventAcronym{CVIT}
\EventYear{2016}
\EventDate{December 24--27, 2016}
\EventLocation{Little Whinging, United Kingdom}
\EventLogo{}
\SeriesVolume{42}
\ArticleNo{23}

\begin{document}

\maketitle

\begin{abstract}
Given a graph having $n$ vertices and up to $m$ edges, we consider the maintenance of its Maximal Independent Set (MIS), subject to an online sequence of $u$ edge insertions and/or deletions in the prediction model. The algorithm \textit{leverages} a predicted sequence of updates, which can be preprocessed to aid during the online sequence of updates. We study the problem in the incremental (insertion-only), decremental (deletions-only), and fully dynamic (insertions and deletions) settings for edge updates. The state-of-the-art non-prediction deterministic algorithms for decremental, incremental, and fully dynamic settings respectively require amortized $O(1),O(\sqrt{m})$ and $O(\min\{n,m^{2/3}\})$ update time, which can be $O(n),O(m)$ and $O(m)$ in the worst case. Recently, Bernstein et al.~[SODA26] showed that for combinatorial algorithms (even randomized with an adaptive adversary), incremental (and hence fully dynamic) algorithms  require $\Omega(n^{1+o(1)})$ amortized update time using the combinatorial Boolean Matrix Multiplication (BMM) conjecture.

We describe algorithms whose complexities are expressed as error measures of the predicted sequence. We define $\eta_e$ as the length of the suffix of the predicted edge sequence that is incorrect, and $\mathbf{n}_e$ as the number of vertices on which such edges are incident. For fully dynamic updates, we consider $\eta_e$ removing duplicates and $\eta'_e$ keeping duplicates.
We define $\eta_v$ as the number of vertices whose MIS status was incorrectly predicted due to the error in the predicted update sequence, $\eta^*_v$ as the sum of degrees of such vertices, and $\mathbf{n}_v$ as the number of neighbours of such vertices. Note that by definition $\eta_e,\eta^*_v\leq 2m$, $\mathbf{n}_e,\mathbf{n}_v,\eta_v\le n$, $\eta'_e\leq u$, and $\mathbf{n}_e\leq 2\eta_e,\mathbf{n}_v\leq \eta^*_v$.

\begin{enumerate}
    \item \textit{Decremental MIS.} Our algorithm requires $O(\eta_v)$ worst-case update time, using $O(mn)$  preprocessing time and space, requiring total $O(m+\eta_v\eta_e)$ update time. The latter can be improved to  $O(m)$ space and preprocessing time with $O(m+\mathbf{n}_e\eta_v+n\log \frac{m}{n})$ total update time at the expense of increased $O(\log n+\eta_v)$ worst case update time. 
    \item \textit{Incremental MIS.} Our algorithm requires $O(\eta_e+\eta^*_v)$ worst-case update time, using $O(mn)$ preprocessing time and space, requiring total   $O(m+\eta^2_e+\eta^*_v\eta_e)$ update time. The latter can be improved to $O(m\sqrt{m})$ space and preprocessing time with $O(m+\eta_e\sqrt{\eta_e}+\eta_e\eta^*_v+n\log \frac{m}{n})$ total update time at the expense of increased $O(\log n+\mathbf{n}_v\log n+\eta_e + \eta_v^*)$ worst-case update time. 
    \item \textit{Fully dynamic MIS.} Our algorithm requires $O(\eta_e+\eta^*_v)$ worst-case update time, using $O(un)$ preprocessing time and space, requiring total   $O(u+\eta'_e\eta_e+\eta'_e\eta^*_v)$ update time.
\end{enumerate}
At its core our algorithms use batch update versions of the MIS problem. This allows us to use the error correction framework of Khan et al.~[Arxiv26], which improves the error measures to Prefix Set matching for partially dynamic and Set Edit Distance for fully dynamic problems, maintaining the same bounds. 
\end{abstract}

\newpage
\setcounter{page}{1}
\section{Introduction}
\label{sec:Introduction}

Dynamic graph algorithms study the maintenance of graph-theoretic structures
as the underlying graph is modified by an online sequence of edge insertions and/or
deletions. The goal is to repair the maintained solution after every update
much faster than recomputing it from scratch. An algorithm is termed \emph{fully dynamic} if it supports both insertions and deletions, and \emph{partially dynamic}, specifically \emph{incremental} or \emph{decremental}, when the update sequence is restricted to only insertions or only deletions, respectively. A \emph{fault tolerant} algorithm repairs the solution after a set of $k$ edge failures. 
Prominent problems studied in this domain include connectivity~\cite{HolmLT01,KapronKM13,ChuzhoyGLNPS20}, reachability~\cite{even-shiloach,HenzingerKN14,DemetrescuI08}, and shortest
paths~\cite{DemetrescuI04,Thorup05,RodittyZ11}, where most results are proven conditionally optimal under popular conjectures~\cite{RodittyZ11,AbboudW14,HenzingerKNS15}.

We investigate the maintenance of a \emph{Maximal Independent Set (MIS)} under edge
updates. Given a graph $G(V,E)$ with $|V|=n$ and $|E|=m$, a set
$\mathcal{M}\subseteq V$ is an MIS if (i) no two vertices in $\mathcal{M}$ are adjacent
(\emph{independence}), and (ii) every vertex outside $\mathcal{M}$ has at least one
neighbour inside it (\emph{maximality}). Although computing a maximum cardinality independent set is NP-hard~\cite{GareyJ90}, a simple greedy algorithm computes a maximal independent set (MIS) in $O(m+n)$ time. 
In the dynamic setting, MIS can thus be trivially maintained in $O(m+n)$ update time by recomputing it from scratch after each update. Censor-Hillel et al.~\cite{CensorHK16} presented the first non-trivial algorithm requiring $O(n)$ amortized update time. Assadi et al.~\cite{AssadiOSS18} improved it to an $O(m^{3/4})$ time algorithm, which was later improved to $O(m^{2/3})$~\cite{GuptaK21,GuptaKS26}. They also showed that decremental and incremental MIS can be maintained using $O(1)$ and $O(\sqrt{m})$ amortized update times, respectively. However, the worst-case update time was $O(n)$ for decremental and $O(m)$ for incremental settings. 
Allowing randomization, even the fully dynamic MIS can be updated in $O(poly \log n)$~\cite{ChechikZ19,BehnezhadDHSS19} time, assuming an oblivious adversary. Recently, Bernstein et al.~\cite{BernsteinBFKS26} proved assuming combinatorial Boolean Matrix Multiplication (BMM) conjecture, even a randomized incremental MIS  algorithm with an adaptive adversary requires $\Omega(n^{1-o(1)})$ amortized update time. 

In this paper, we study the dynamic MIS problem in the \emph{prediction model}: at the beginning of the computation, the algorithm is given
a predicted update sequence $\hat{U}$ (e.g., produced by a learned oracle), which can be
preprocessed offline; the real sequence $U$ is then revealed online, one update at a
time, and after each update the data structure must report a valid MIS of the current
graph.
The motivation behind the prediction model is that the real-world update sequences are
rarely adversarial and tend to follow patterns due to often repetitive user behaviours. As a result, future updates can be anticipated from historical data or learned models. This emerging area of algorithm design is called \emph{algorithms
with predictions} (or \emph{learning-augmented algorithms}), which formalizes
this intuition by equipping classical algorithms with a \emph{prediction}—an
auxiliary input, potentially produced by a machine-learning oracle. The goal is to develop provably faster algorithms while retaining correctness guarantees regardless of prediction quality. Such algorithms are expected to demonstrate  \emph{consistency} (optimal performance for perfect predictions),
\emph{robustness} (comparable performance to the best prediction-free algorithm even for worst prediction), and
\emph{smooth degradation} (graceful decrease in performance between the two
extremes). Recently, the prediction algorithms have received significant attention, notably for online problems (survey in~\cite{MV21,MV22}) as well as for dynamic graph algorithms such as dynamic all pairs shortest paths~\cite{brand2024dynamic}, breadth first search~\cite{KhanKL26}, incremental topological ordering and cycle detection~\cite{mccauley2024incremental}. Liu and Srinivas~\cite{liu2024predicted} presented a general framework applicable to a plethora of problems, while Henzinger et al.~\cite{henzinger2024complexity} focused on conditional lower bounds. Notably, these results employed different \textit{error measures} to analyze their algorithms, typically fine-tuned for the studied problem. 

\paragraph*{Our Results}
We present the first algorithms for dynamic MIS in the prediction model, for
incremental, decremental, and fully dynamic settings. 
Similar to previous dynamic graph algorithms with predictions, our complexity bounds are expressed in terms of prediction-quality error measures related to the maintained MIS. We define $\eta_e$ as the length of the suffix of the predicted edge sequence that is incorrect, and $\mathbf{n}_e$ as the number of vertices on which such edges are incident. For fully dynamic updates we consider $\eta_e$ removing duplicates, and $\eta'_e$ keeping duplicates. We define $\eta_v$ as the number of vertices whose MIS status was incorrectly predicted with respect to the current output MIS, $\eta^*_v$ as the sum of degrees of such vertices, and $\mathbf{n}_v$ as the number of vertices that are the neighbours of such vertices. Note that $\eta_e,\eta^*_v\leq 2m$, $\eta_v,\mathbf{n}_e,\mathbf{n}_v\le n$ and $\eta'_e\leq u$. Also, since multiple MIS may exist, the values of $\eta_v$ and $\eta^*_v$ may vary based on the maintained MIS. However, since it captures the difference of predicted MIS (which is invalid) with respect to some valid MIS, it captures the amount of local repair a valid algorithm may perform, signifying the impact of prediction error. 

\begin{enumerate}
\item Prediction algorithms for Decremental MIS
\begin{restatable}[Decremental MIS]{theorem}{decMIS}
Given a graph with $n$ vertices and $m$ edges under deletion of edges, its MIS can be maintained using $O(mn)$ preprocessing time and space, with worst case update time  $O(\eta_v)$ and total update time $O(m+\eta_e\eta_v)$. 
\label{theorem:naiveDecremental} 
\end{restatable}
\begin{restatable}[Improved Decremental MIS]{theorem}{decMISimp}
Given a graph with $n$ vertices and $m$ edges under deletion of edges, its MIS can be maintained using $O(m)$ preprocessing time and space, with worst case  $O(\eta_v+\log n)$ and total $O(m+\eta_v \mathbf{n}_e + n \log \frac{m}{n})$ update times.
\label{theorem:impDecremental} 
\end{restatable}

\item Prediction algorithms for Incremental MIS

\begin{restatable}[Incremental MIS]{theorem}{incMIS}
Given an empty graph with $n$ vertices under insertion of $m$ edges, its MIS can be maintained using $O(mn)$ preprocessing time and space, with worst-case update time $O(\eta_e + \eta_v^*)$ and total update time $O(m + \eta^2_e +\eta_e\eta_v^*)$.
\label{theorem:Incremental} 
\end{restatable}

\begin{restatable}[Improved Incremental MIS]{theorem}{incMISimp}
Given an empty graph with $n$ vertices under insertion of $m$ edges, its MIS can be maintained using $O(m\sqrt{m})$ preprocessing time and space, with worst-case update time $O(\log n+\mathbf{n}_v\log n+\eta_e + \eta_v^*)$ and total update time $O(m+\eta_e\sqrt{\eta_e} +\eta_e\eta_v^*+n\log \frac{m}{n})$.
\label{theorem:Incrementalimp} 
\end{restatable}

\item Prediction algorithms for Fully dynamic MIS

\begin{restatable}[Fully Dynamic  MIS]{theorem}{fdMIS}
Given a graph with $n$ vertices under $u$ edge updates, its MIS can be maintained using $O(un)$ preprocessing time and space, with worst-case update time $O(\eta_e + \eta_v^*)$ and total update time $O(u+\eta'_e\eta_e+\eta'_e\eta_v^*)$.
\label{theorem:fd} 
\end{restatable}
\end{enumerate}

\begin{remark}
Above bounds are for imperfect predictions, while for perfect predictions our algorithm requires optimal time. This satisfies consistency, robustness, and smooth degradation of prediction algorithms as worst case update time of decremental MIS is $O(n)$ (while $\eta_v\leq n)$, whereas that of incremental and fully dynamic MIS is $O(m)$  (while $\eta_e,\eta^*_v\leq 2m$).
\end{remark}

\begin{remark}
The improvement in update time (arriving online) at the expense of preprocessing time (offline) is non-trivial tradeoff to achieve the desired worst case bounds. 
\end{remark}

\begin{remark}
Our error measures starting from the first difference between real and predicted updates are seemingly stringent. However, we can use the generalized error correction framework of Khan et al.~\cite{KhanKL26}, who use a similar first difference for error model. 
\begin{enumerate}
    \item For the partially dynamic case, their framework relaxes the first difference model to Prefix Match (the maximum prefix of the predicted sequence present in the update sequence) using $O(1)$ overhead in update time.
    \item For fully dynamic case, their framework relaxes the first difference model to Set Edit Distance (minimum set difference from any point in predicted sequence with current update sequence) using $O(\eta_e)$ overhead in update time.
\end{enumerate}
  Thus, our bounds extend to the relaxed error model of Prefix Match (partially dynamic) and Set Edit Distance (fully dynamic).
\end{remark}

\section{Preliminary}
Given a graph $G(V,E)$ having $n$ vertices undergoing $u$ edge updates, with a maximum of $m$ edges during updates. Assume a sufficiently dense graph with $m\geq n-1$. We denote the initial graph as  $G_0=(V,E_0)$ with an edge set $E_0\subseteq V\times V$, where $m_0=|E_0|$. The graph
is subject to an online sequence of $u$ edge updates 
$U=\{\langle e_1,t_1\rangle,\langle e_2,t_2 \rangle,\dots,\langle e_u,t_u\rangle\}$, where each $e_i$ is an edge which is inserted ($t_i=+$)
or deleted ($t_i=-$). In case of incremental and decremental ${t_i}$ is known ($+$ or $-$), so we omit ${t_i}$ completely. For partially dynamic cases, we assume $u=m$ starting or ending with an empty graph. However, for the fully dynamic case $u$ can potentially be much larger than $m$. 
We denote by $G_i=(V,E_i)$ the graph obtained
after applying updates $\{\langle e_1,t_1\rangle,\cdots,\langle e_i,t_i \rangle\}$ to $G_0$, where $m_i=|E_i|$. When the current graph $G_k(V,E_k)$ is clear from the context, we simply use $G(V,E)$ having $m$ edges. 
For a vertex $v \in V$, $N(v)$ denotes the set of neighbors of $v$ in $G_k$, i.e.,
$N(v) = \{u \in V \mid (u,v) \in E_k\}$.
We denote the degree of $v$ by $\deg(v) = |N(v)|$. While considering $\deg(v)$ over multiple graphs (or updates), we simply assume it to be the maximum over all the graphs. 
For a set of edges $F$, $V(F)$ denotes the set of vertices on which the edges are incident.
We denote the MIS of the current graph $G_k$ as ${\cal M}_k$ or simply $\cal M$.  We also denote $count_k[v]$ (or simply $count[v]$) as the number of neighbours of $v$ in $\cal M$, or $|N(v)\cap {\cal M}|$.  
Note that by definition, a vertex $v$ belongs to $\cal M$ if and only if $count[v] = 0$.

\paragraph*{Prediction Model for Dynamic Updates}
We study the dynamic MIS problem under a prediction model. Let us assume the predicted edge update sequence is 
$\hat{U}=\{\langle \hat{e}_1,\hat{t}_1\rangle,\langle\hat{e}_2, \hat{t}_2\rangle ,\cdots,\langle\hat{e}_u,\hat{t}_u\rangle \}$,
where $\hat{t}_i$ can be $+$ or $-$ representing insertion or deletion, respectively. Again, for partially dynamic, we assume $u=m$ and ignore $\hat{t}_i$. Also, the graph after applying updates $\hat{U}_i=\{\langle \hat{e}_1,\hat{t}_1\rangle,\cdots,\langle\hat{e}_i,\hat{t}_i\rangle \}$ is represented by $\hat{G}_i$.
Assume that currently only $k$ edges are updated in the actual sequence
$U=\{\langle e_1,t_1\rangle, \langle e_2,t_2\rangle, \cdots,\langle e_k,t_k\rangle\}.$
Let $i^*$ be the maximum number of updates starting from $\langle e_1,t_1\rangle$ that exactly match between the predicted and actual update sequences, i.e., 
$\langle \hat{e}_i,\hat{t}_i\rangle = \langle e_i,t_i\rangle \quad \text{for } i \leq i^*,$ and $\langle \hat{e}_{i^*+1},\hat{t}_{i^*+1}\rangle \neq \langle e_{i^*+1},t_{i^*+1}\rangle$. We define the following:

\begin{definition}[Edge Prediction Error Measures $E_{err},\eta_e,\eta'_e,\mathbf{n}_e$]
We define the following error measures related to ${G_k}$ with respect to $\hat{G}_{i^*}$:
\begin{enumerate}
    \item \textbf{Error Edges ($E_{err}$).}
The set of edges updates since the first prediction error occurred, i.e., $E_{err}=\{e_{i^*+1},\cdots,e_{k}\}$.    

\item \textbf{Prediction Error ($\eta_e,\eta'_e$).}
The number of updates since the first prediction error occurred, i.e., $\eta_e=|E_{err}|$ (without duplicates) and  $\eta'_e=k-i^*$ (allowing duplicates).

\item \textbf{Error Incidence ($\mathbf{n}_e$).}
The number of vertices on which the edges are incident, since the first prediction error occurred, i.e., $\mathbf{n}_e=|V(E_{err})|$.    
\end{enumerate}
\end{definition}

\newpage

\begin{definition}[MIS Prediction Error Measures $V_{err},\eta_v,\eta^*_v,\mathbf{n}_v$]
We define the following error measures related to MIS of $G_i$ maximized over $(i^*<i\leq k)$ with respect to that of $\hat{G}_{i^*}$:
\begin{enumerate}
\item \textbf{Error Vertices ($V_{err}$).} The set of vertices whose membership in $\cal M$ changed after the first prediction error has occurred, i.e., $V_{err}=({\cal M}_i-{\cal M}_{i^*})\cup ({\cal M}_{i^*}-{\cal M}_i)$.

\item \textbf{Prediction Error ($\eta_v$).}
The number of vertices whose membership in $\cal M$ changes after the first prediction error has occurred, i.e., $\eta_v=|V_{err}|$.

\item \textbf{Error Incidence ($\mathbf{n}_v$).}
The number of distinct neighbours of the vertices whose membership in $\cal M$ changed after the first prediction error has occurred, i.e., $\mathbf{n}_v=|\bigcup_{v\in V_{err}} N(v)|$.    

\item \textbf{Weighted Prediction Error ($\eta^*_v$).}
The sum of degrees of vertices whose membership in $\cal M$ changed after the first prediction error has occurred, i.e., $\eta^*_v=\sum_{v\in V_{err}} \deg(v)$. 
\end{enumerate}
\end{definition}

\begin{remark}
    Note that edge prediction error measures $\eta_e,\eta'_e,\mathbf{n}_e$ are non-decreasing with $k$ starting from the first error $i^*+1$. However, the MIS prediction error measures may decrease with increasing $k$. Hence for the sake of simplicity we consider $\eta_v,\eta^*_v$ to be the maximum such error for $G_i$ for $i^*<i\leq k$.
\end{remark}

\begin{remark}
    By definition $\eta_e,\eta^*_v\leq 2m$, $\mathbf{n}_e,\mathbf{n}_v,\eta_v\le n$ and $\eta'_e\leq u$. Also, $\mathbf{n}_e\leq 2\eta_e,\mathbf{n}_v\leq \eta^*_v$. 
\end{remark}
\begin{remark}
Since multiple MIS exist for a graph $G$, the measure of $\eta_v$ and $\eta^*_v$ depends on the maintained MIS $\cal M$ and not just $G$. However, since it captures the difference of predicted MIS (which is invalid) with respect to some valid MIS, it measures the amount of local repair a valid algorithm may perform, signifying the impact of the prediction error. 
    
\end{remark}

\paragraph*{State of the art dynamic MIS}
We shall use the following results about dynamic MIS for comparison or as a black box. 
\begin{theorem}[Simple Dynamic MIS~\cite{AssadiOSS18,GuptaKS26}]
Given a graph having $n$ vertices and $m$ edges under edge updates, using $O(m)$ space and preprocessing time its 
    \begin{enumerate}
        \item[(a)] fully dynamic MIS is maintained in $O(n)$ amortized and $O(m)$ worst case update time,
        \item [(b)] incremental MIS is maintained in $O(\sqrt{m})$ amortized and $O(m)$ worst case update time,
        \item[(c)] decremental MIS is maintained in $O(1)$ amortized and $O(n)$ worst case update time,
    \end{enumerate}
     \textit{explicitly} storing count of MIS neighbours, with $O(1)$ amortized adjustments to the MIS.
    \label{thm:dynMIS}
\end{theorem}

\begin{theorem}[Improved Dynamic MIS~\cite{GuptaK21,GuptaKS26}]
    Given a graph having $n$ vertices and $m$ edges, using $O(m)$ space and preprocessing time its dynamic MIS can be maintained under
    \begin{enumerate}
        \item[(a)] fully dynamic edge updates in $O(m^{2/3})$ amortized and $O(m)$ worst case update time.
        \item[(b)] fault tolerant updates in $O(k+\mathbf{n}_k^{2})$ time, for $k$ edge failures incident on  $\mathbf{n_k}$ vertices. 
    \end{enumerate}
\label{thm:dynMIS2}
\end{theorem}

We also use the following additional property of the decremental MIS algorithm~\cite{AssadiOSS18}.
\begin{lemma}
    \label{lem:decMIS}
At most one vertex can be added to $\cal M$ and none removed during an update. 
\end{lemma}

In our paper, we require an extension of incremental MIS algorithms for edge updates to also allow newly inserted edge to be incident on one or two new vertices. If the inserted edge $(u,v)$ has $v$ as only new vertex, we can simply assign it to be in MIS based on the MIS status of $u$ and update the counts in $O(1)$ time. If the inserted edge $(u,v)$ has both $u$ and $v$ as new vertices, we can simply add one of them to MIS and update the count of other in $O(1)$ time. Hence we have the following result.

\begin{corollary}[Extended Incremental MIS]
\label{col:extInc}
    Given a graph having $n$ vertices and $m$ edges, using $O(m)$ space and preprocessing time its MIS can be maintained under incremental edge updates where new edges are possibly incident on new vertices in $O(\sqrt{m})$ amortized and $O(m)$ worst case update time, explicitly storing count of MIS neighbours, with $O(1)$ amortized adjustments to the MIS.
\end{corollary}

\newpage
\section{Decremental MIS with Prediction}
We demonstrate how predicted update sequence is leveraged to improve the worst case update time of decremental MIS from $O(n)$ to $O(\eta_v)$, using $O(mn)$ preprocessing time and space.

\subsection{Data Structure}
We store a count array $C[m][n]$, where $C[i][v]$ represents the count of MIS neighbors of $v$ in $\hat{G}_i$.
We also maintain $\hat{\cal M}^+_i$ storing the list of vertices inserted to $\hat{\cal M}_{i-1}$ to get $\hat{\cal M}_i$. 
After the first incorrect prediction $\hat{e}_{i^*+1}\neq e_{i^*+1}$, we maintain a list of vertices ${\cal M}^+$ added to MIS. Thereafter, $C[i^*][v]$ represents the count of MIS neighbours of $v\in V$ except for those added to ${\cal M}^+$ (i.e., only ${\cal M}_{i^*}$), possibly violating $v\in \cal M$ iff $C[i^*][v]=0$. We also explicitly maintain $\cal M$ as the  MIS of the current graph.

\subsection{Algorithm}
We avoid the expensive operation of updating counts of neighbours of a vertex entering $\cal M$. Instead we maintain a list of such vertices ${\cal M}^+$ and process them explicitly when required. Similar technique was used in the fault tolerant MIS algorithm (see \Cref{thm:dynMIS2}).\\

\noindent
\textbf{Preprocessing.} We use the decremental MIS algorithm (\Cref{thm:dynMIS}) to process the predicted update sequence $\hat{U}$ maintaining the count of MIS neighbours for each vertex. After each update $\hat{e}_i$, we copy the counts of each vertex $v\in V$ in $C[i][v]$. Also, the corresponding changes from $\hat{\cal M}_{i-1}$ to $\hat{\cal M}_{i}$ are updated in $\hat{\cal M}^+_i$. Recall that the algorithm only adds vertices to MIS and never removes them (\Cref{lem:decMIS}). We also initialize ${\cal M}^+$ to $\emptyset$.\\

\noindent
\textbf{Update.}
While the update $e_i=\hat{e}_i$, we only update $\cal M$ using $\hat{\cal M}^+_i$. After the last such edge $e_{i^*}=\hat{e}_{i^*}$, where $e_{i^*+1}\neq \hat{e}_{i^*+1}$, we continue to use the last correctly predicted count $C[i^*][\cdot]$ for future processing. 
Now, on deletion of an edge $(u,v)$ (see \Cref{alg:delete}), we need to update only when one of $u$ or $v$ (say $u$) is in $\cal M$. Clearly, due to independence $v\notin\cal M$. However, this reduces $C[i^*][v]$ (if $u\in {\cal M}_{i^*}$) which can become zero and thus potentially be added to $\cal M$ and ${\cal M}^+$. Since on adding to ${\cal M}^+$ a vertex does not update the count of its neighbours, it is possible $v$ has neighbours in ${\cal M}^+$. We thus verify if it has any neighbours in ${\cal M}^+$, in the absence of  which $v$ is added to ${\cal M}^+$ and $\cal M$.

\begin{algorithm}
\caption{$Delete(u,v)$: Update MIS after deletion of edge $(u,v)$ post prediction error.}
\label{alg:delete}
\begin{multicols}{2}
\BlankLine
Remove $(u,v)$ from $G$\;
\lIf{$v\in \cal M$} 
    {Exchange $u$ and $v$}
\lIf{$u\notin \cal M$}{
    {\textbf{return}}
}
\BlankLine
\tcc{Hence $u\in \cal M$ and $v\notin \cal M$}
\If{$u\in \cal M$ and $u\notin {\cal M}^+$}{
$C[i^*][v] \gets C[i^*][v]-1$\;
}
\If{$C[i^*][v]==0$}{
    \ForAll{$x \in {\cal M}^+$} {
    \lIf{$x\in N(v)$} {
        \textbf{return}
    }
}
Add $v$ to ${\cal M},{\cal M}^+$\;
}
\end{multicols}
\end{algorithm}

\nolinenumbers
\begin{figure}[!h]
\centering
\subcaptionbox{Initial state
}[0.30\linewidth]{%
\begin{tikzpicture}[scale=0.72, transform shape,
    every node/.style={circle, draw, minimum size=6mm, inner sep=0pt},
    dark/.style={fill=cyan!40, text=black},
    light/.style={fill=gray!20, text=black},
    thick
]
\node[dark,fill=gray!20, label=above:{\scriptsize $1$}] (a)  at (0,3) {a};
\node[light, fill=cyan!40, label=above:{\scriptsize $0$}] (b) at (1.3,3.5) {b};
\node[light, label=above:{\scriptsize $1$}] (i) at (2.7,3) {i};
\node[light, label=above:{\scriptsize $1$}] (j) at (3.7,3.5) {j};

\node[dark,fill=cyan!40, , label=above:{\scriptsize $0$}] (c)  at (1,1.5) {c};
\node[light, label=above:{\scriptsize $1$}] (d) at (2.5,1) {d};
\node[light, label=above:{\scriptsize $1$}] (e) at (0.5,0.5) {e};
\node[light, label=above:{\scriptsize $1$}] (l) at (3.7,1) {l};
\node[dark, fill=cyan!40, label=above:{\scriptsize $0$}] (m) at (4.4,2) {m};

\draw (a) -- (b);
\draw (b) -- (i);
\draw (i) -- (j);

\draw (c) -- (d);
\draw (d) -- (e);
\draw (e) -- (c);
\draw (d) -- (l);
\draw (j) -- (m);
\draw (l) -- (m);
\draw (i) to[bend left=15] (d);
\draw (i) -- (l);

\end{tikzpicture}
}%
\subcaptionbox{After deleting ${(b,i),(c,d),(c,e)}$
}[0.30\linewidth]{%
\begin{tikzpicture}[scale=0.72, transform shape,
    every node/.style={circle, draw, minimum size=6mm, inner sep=0pt},
    dark/.style={fill=cyan!40, text=black},
    light/.style={fill=gray!20, text=black},
    thick
]
\node[dark,fill=gray!20, label=above:{\scriptsize $1$}] (a)  at (0,3) {a};
\node[light, fill=cyan!40, label=above:{\scriptsize $0$}] (b) at (1.3,3.5) {b};
\node[light, fill=yellow!40, label=above:{\scriptsize $0$}] (i) at (2.7,3) {i};
\node[light, label=above:{\scriptsize $1$}] (j) at (3.7,3.5) {j};

\node[dark,fill=cyan!40, , label=above:{\scriptsize $0$}] (c)  at (1,1.5) {c};
\node[light, label=above:{\scriptsize $0$}] (d) at (2.5,1) {d};
\node[light, fill=yellow!40, label=above:{\scriptsize $0$}] (e) at (0.5,0.5) {e};
\node[light, label=above:{\scriptsize $1$}] (l) at (3.7,1) {l};
\node[dark, fill=cyan!40, label=above:{\scriptsize $0$}] (m) at (4.4,2) {m};

\draw (a) -- (b);
\draw (i) -- (j);

\draw (d) -- (e);
\draw (d) -- (l);
\draw (l) -- (m);
\draw (j) -- (m);
\draw (i) to[bend left=15] (d);
\draw (i) -- (l);

\end{tikzpicture}
}
\subcaptionbox{After deleting $(d,i),(d,e)$}[0.30\linewidth]{%
\begin{tikzpicture}[scale=0.72, transform shape,
    every node/.style={circle, draw, minimum size=6mm, inner sep=0pt},
    dark/.style={fill=cyan!40, text=black},
    light/.style={fill=gray!20, text=black},
    thick
]
\node[dark,fill=gray!20, label=above:{\scriptsize $1$}] (a)  at (0,3) {a};
\node[light, fill=cyan!40, label=above:{\scriptsize $0$}] (b) at (1.3,3.5) {b};
\node[light,fill=yellow!40, label=above:{\scriptsize $0$}] (i) at (2.7,3) {i};
\node[light, label=above:{\scriptsize $1$}] (j) at (3.7,3.5) {j};

\node[dark,fill=cyan!40, , label=above:{\scriptsize $0$}] (c)  at (1,1.5) {c};
\node[light,fill=yellow!40, label=above:{\scriptsize $0$}] (d) at (2.5,1) {d};
\node[light, fill=yellow!40, label=above:{\scriptsize $0$}] (e) at (0.5,0.5) {e};
\node[light, label=above:{\scriptsize $1$}] (l) at (3.7,1) {l};
\node[dark, fill=cyan!40, label=above:{\scriptsize $0$}] (m) at (4.4,2) {m};

\draw (a) -- (b);
\draw (i) -- (j);

\draw (d) -- (l);
\draw (l) -- (m);
\draw (j) -- (m);
\draw (i) -- (l);

\end{tikzpicture}
}

\caption{Example where ${\cal M}^+$ are shown in yellow, and remaining $\cal M$ in blue.}
\label{fig:dec-eg}
\end{figure}
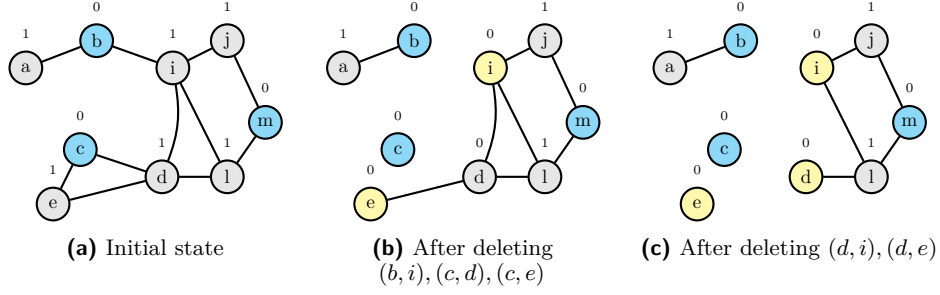

Consider \Cref{fig:dec-eg}, initially after $i^*$ updates, the $C[i^*][\cdot]$ is shown for each vertex with ${\cal M}_{i^*}$ shown in blue. On deleting $(b,i)$ count of $i$ reduces to zero, adding it to ${\cal M}^+$. Further, on deleting $(c,d)$ the count of $d$ becomes zero, but after processing $i\in {\cal M}^+$ it remains $0$ but is not added to $\cal M^+$. Now, on deleting $(c,e)$ the count of $e$ becomes zero and is added to ${\cal M}^+$. Now, on deleting $(d,i)$ count of $d$ is again zero, but on processing $e\in {\cal M}^+$ it is still not added to ${\cal M}^+$. Finally, on deleting $(d,e)$ with no neighbours in ${\cal M}^+$, $d$ is added to ${\cal M}^+$.

\subsection{Correctness}
The correctness depends on the invariant: vertices $v\in V$ with $C[i^*][v]=0$ and $N(v)\cap{{\cal M}^+}=\emptyset$ iff $v\in \cal M$.
Independence is ensured since a vertex gets $C[i^*][v]==0$ only when it has not ${\cal M}\setminus{\cal M}^+$ neighbours, and is added to $\cal M$ only after verifying no neighbour exists in ${\cal M}^+$. To ensure maximality, we check the only vertex $v$ which can enter the MIS on deletion of $(u,v)$ with $u\in \cal M$.  If it does not violate the independence it is added to $\cal M$.

\subsection{Analysis}
The preprocessing requires $O(mn)$ time and space to copy $C[i][v]$ for each vertex $v$ corresponding to every edge $\hat{e}_i$, and $O(n)$ to compute $\hat{\cal M}^+_i$ (\Cref{lem:decMIS}), while the simple decremental algorithm takes only $O(m)$ total time. While the prediction is true, each update is processed in $O(1)$ time (recall \Cref{lem:decMIS}) requiring total $O(m)$ update time. Thereafter, for each of the $\eta_e$ edges, the update time is $O(1)$ except processing ${\cal M}^+$, which is $O(\eta_v)$ by definition. This results in total $O(m+\eta_e\eta_v)$ update time. 

\decMIS*
\begin{remark} The above bound is for imperfect prediction only, while for perfect prediction we have optimal bound. The algorithm satisfies the desired properties of consistency, robustness and smooth degradation for the worst case update time since $\eta_v\leq n$. However, it comes at the expense of significantly higher $O(mn)$ space and total update time compared to the $O(m)$ space and total time of the simple decremental MIS algorithm.  \end{remark}


\section{Improved Decremental MIS Algorithm}
\label{sec:impDec}
In this section, we improve our decremental algorithm using a compact data structure (instead of $C[m][n]$) to require $O(m)$  preprocessing time and space at the expense of higher $O(\eta_v+\log n)$ worst case update time. We also avoid repeated processing of ${\cal M}^+$ for the same vertex, reducing the total update time to $O(m+\eta_v\mathbf{n}_e+n\log\frac{m}{n})$.

\subsection{Data Structure}
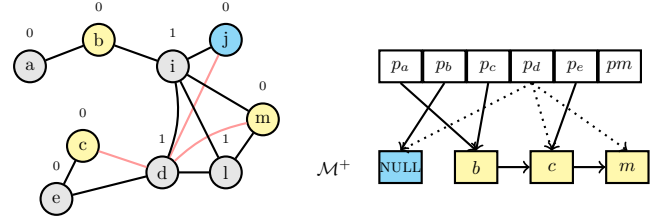
\begin{figure}[h]
\centering
\begin{tikzpicture}[scale=0.7, transform shape,
    every node/.style={circle, draw, minimum size=6mm, inner sep=0pt},
    dark/.style={fill=cyan!40, text=black},
    light/.style={fill=gray!20, text=black},
    yellow/.style={fill=yellow!40, text=black},
    thick,
    box/.style={draw,rectangle, minimum width=8mm, minimum height=6mm}
]
\node[dark,fill=gray!20, label=above:{\scriptsize $0$}] (a)  at (0,3) {a};
\node[light, fill=yellow!40, label=above:{\scriptsize $0$}] (b) at (1.3,3.5) {b};
\node[light, label=above:{\scriptsize $1$}] (i) at (2.7,3) {i};
\node[light, fill=cyan!40, label=above:{\scriptsize $0$}] (j) at (3.7,3.5) {j};

\node[dark, fill=yellow!40 , label=above:{\scriptsize $0$}] (c)  at (1,1.5) {c};
\node[light, label=above:{\scriptsize $1$}] (d) at (2.5,1) {d};
\node[light, label=above:{\scriptsize $0$}] (e) at (0.5,0.5) {e};
\node[light, label=above:{\scriptsize $1$}] (l) at (3.7,1) {l};
\node[dark, fill=yellow!40, label=above:{\scriptsize $0$}] (m) at (4.4,2) {m};

\draw (a) -- (b);
\draw (b) -- (i);
\draw (i) -- (j);

\draw[red!40] (c) -- (d);
\draw (d) -- (e);
\draw[red!40] (d) -- (j);
\draw (e) -- (c);
\draw (d) -- (l);
\draw (d)[red!40] to[bend left=15] (m);
\draw (l) -- (m);
\draw (i) -- (m);
\draw (i) to[bend left=15] (d);
\draw (i) -- (l);

\node[box] (a0) at (7,3.0) {$p_a$};
\node[box, right=0mm of a0] (a1) {$p_b$};
\node[box, right=0mm of a1] (a2) {$p_c$};
\node[box, right=0mm of a2] (a3) {$p_d$};
\node[box, right=0mm of a3] (a4) {$p_e$};
\node[box, right=0mm of a4] (a5) {$pm$};


\node[box, fill=cyan!40] (b0) at (7,1.1) {\textsc{null}};
\node[box, fill=yellow!40, right=6mm of b0] (b1) {$b$};
\node[box, fill=yellow!40, right=6mm of b1] (b2) {$c$};
\node[box, fill=yellow!40,right=6mm of b2] (b3) {$m$};
\draw[->, thick] (b1.east) -- (b2.west);
\draw[->, thick] (b2.east) -- (b3.west);

\node[left=5mm of b0, draw=none, shape=rectangle] {$\mathcal{M^+}$};

\draw[->] (a0.south) -- (b1.north);
\draw[->] (a1.south) -- (b0.north);
\draw[->] (a2.south) -- (b1.north);
\draw[->][dotted] (a3.south) -- (b0.north);
\draw[->][dotted] (a3.south) -- (b2.north);
\draw[->][dotted] (a3.south) -- (b3.north);
\draw[->] (a4.south) -- (b2.north);

\end{tikzpicture}
\caption{Dirty nodes $b,c,m$ added to ${\cal M}^+$ list with vertex pointers to ${\cal M}^+$, also showing counts. Pointer $p_d$ is \textsc{null} initially, on deleting $(d,j)$ processes $b$ then $c$, stopping at $c$ as $c\in N(d)$. Now, on deleting $(d,m)$ again count is $0$, so $d$ processed again starting from $c$ which remains at $c$. Finally, deleting $(d,c)$ it moves to end of the list $m$ but on finding no neighbours, hence $d$ is added to $\cal M^+$.
}
    \label{fig:dirty}
\end{figure}

We store the counts compactly in an array $C'[n]$ of dynamic arrays. If count of MIS neighbours of $v$ changed to $c$ from $\hat{G}_{i-1}$ to $\hat{G}_{i}$, we add the pair $(i,c)$ to $C'[v]$. Note that each $C'[v]$ is sorted according to the corresponding $i$. 
We also maintain $\hat{\cal M}^+_i$ storing the list of vertices inserted in to $\hat{\cal M}_{i-1}$ to get $\hat{\cal M}_i$ (\Cref{lem:decMIS}).  
Again, after the first incorrect prediction $\hat{e}_{i^*+1}\neq e_{i^*+1}$, we maintain a list of vertices ${\cal M}^+$ added to MIS. Thereafter, $C'[v]$'s entry $(i,c_v)$ with last $i\leq i^*$ stores the count of MIS neighbours of $v\in V$ except for those added to ${\cal M}^+$ (i.e. only ${\cal M}_{i^*}$), maintaining $v\in \cal M$ iff $c_v==0$. The corresponding index in $C'[v]$ of the entry is stored in $j_v$. We additionally store for each vertex $v$, a pointer $p_v$ to ${\cal M}^+$ that stores the last element of ${\cal M}^+$ checked for $v$ (see \Cref{fig:dirty}). We also explicitly maintain $\cal M$ as the  MIS of the current graph.

\subsection{Algorithm}
We avoid repeatedly processing ${\cal M}^+$ completely for each $v$ using pointer $p_v$. Also, for using the compact $C'[n]$ instead of $C[m][n]$ (\Cref{alg:delete}), we additionally use binary search.\\

\noindent
\textbf{Preprocessing.} 
We again use the simple decremental MIS algorithm to process the predicted sequence $\hat{U}$, maintaining the count of MIS neighbours of each vertex. On processing $\hat{e}_i$, if the count of $v$ changes to $c$ we add  $(i,c)$ to $C'[v]$, implicitly sorting $C'[v]$. Also, the corresponding changes from $\hat{\cal M}_{i-1}$ to $\hat{\cal M}_{i}$ are updated in $\hat{\cal M}^+_i$ as vertices are only added (\Cref{lem:decMIS}). We also initialize ${\cal M}^+=\emptyset$, and $j_v=-1,p_v=$\textsc{null} for each $v\in V$.\\

\noindent
\textbf{Update.}
While the update $e_i=\hat{e}_i$, we only update $\cal M$ using $\hat{\cal M}^+_i$.
Further, after the last such edge $e_{i^*}=\hat{e}_{i^*}$ where $e_{i^*+1}\neq \hat{e}_{i^*+1}$, we continue to use the counts corresponding to $i^*$. This is computed using binary search once for each vertex $v$, by searching for the highest $(i,c_v)\in C'[v]$ where $i\leq i^*$. Now, when an edge $(u,v)$ is deleted (see \Cref{alg:delete-2}), we follow the same approach, except the vertices in ${\cal M}^+$ are not checked again for neighbours of same vertex $v$. We thus use pointer $p_v$ to check from the last checked vertex in ${\cal M}^+$.

\begin{algorithm}
\caption{$DeleteI(u,v)$: Update MIS after deletion of $(u,v)$ post prediction error.}
\label{alg:delete-2}
\begin{multicols}{2}
Remove $(u,v)$ from $G$\;
\If{$j_u==-1$}{
$j_u\gets$ last $(i,\cdot)\in C'[u]$ with $i\leq i^*$\;}
\If{$j_v==-1$}{
$j_v\gets$ last $(i,.)\in C'[v]$ with $i\leq i^*$\;}
$(i_u,c_u)\gets C'[u][j_u]$\;
$(i_v,c_v)\gets C'[v][j_v]$\;
\BlankLine
\lIf{$v\in \cal M$} 
    {Exchange $u$ and $v$}
\lIf{$u\notin \cal M$}{\textbf{return}}

\tcc{Hence $u\in \cal M$ and $v\notin \cal M$}
\If{$u\in \cal M$ and $u\notin {\cal M}^+$}{
$c_v \gets c_v-1$, update $C'[v][j_v]$\;
}
\BlankLine

\If{$c_v==0$}{
    \lIf{$p_v$ is \textsc{null}}{$p_v\gets $ start of ${\cal M}^{+}$}
    \While{$p_v\neq$ end of ${\cal M}^{+}$} {
    \If{$p_v$ points to $x\in N(v)$} {
       \textbf{return}\;
    }
    $p_v\gets p_v.next$\;
    }
    Add $v$ to ${\cal M},{\cal M}^{+}$\;
}

\end{multicols}
\end{algorithm}

\subsection{Correctness}
The correctness follows from that of \Cref{alg:delete}, since we are only using alternate data structures to access the same information efficiently. Note that the pointers $p_v$ recheck the last checked vertex $x$ as deleted $u$ might be different than $x$.

\subsection{Analysis}
We first analyze the size of $C'[n]$. The count of a vertex $v$ can change due to its neighbour $w$, only when $w$ enters the MIS or $(u,w)$ is deleted (\Cref{lem:decMIS}). Thus, we have $|C'[v]|=O(\deg(v))$, resulting in total $O(m)$ space for $C'[n]$. Now, the preprocessing also requires $O(m)$ time due to the decremental MIS algorithm, thereby computing $C'[n]$ and $\hat{\cal M}^+_i$. Again, while the prediction is true, each update requires $O(1)$ time, resulting in total $O(m)$ time. Thereafter, for the remaining updates in addition to processing ${\cal M}^+$ we require constant binary searches, taking $O(\log n+\eta_v)$ worst case update time. Now, ${\cal M}^+$ will only be processed once in $O(\eta_v)$ for the $O(\mathbf{n}_e)$ vertices on which the deleted edges are incident, taking total $O(\eta_v\mathbf{n}_e)$ time.  Also, the binary search is required only once for each vertex $v$, requiring $O(\log \deg(v))$ time. Using AM-GM inequality, we have the following
$$\sum_{v\in V} \log\deg(v)=\log \prod_{v\in V} \deg(v)=n\log \sqrt[n]{\prod_{v\in V} \deg(v)}\leq n\log \frac{\sum_{v\in V} \deg(v)}{n}=O(n\log \frac{m}{n})$$
 Thus, the total  update time is $O(m+\eta_v\mathbf{n}_e+n\log\frac{m}{n})$.

\decMISimp*
\begin{remark}
The improved algorithm requires linear space while worst case bounds are still robust up to logarithmic factors. However, the total time can still be $O(n^2)$ for worst prediction which is robust only for dense graphs.  
\end{remark}

\section{Incremental MIS with Prediction}
\label{sec:incMIS}
We demonstrate how predicted updates are leveraged to improve the worst case update time of incremental MIS from $O(m)$ to $O(\eta_e+\eta^*_v)$, using $O(mn)$ preprocessing time and space.

\subsection{Data Structure}
We store a count array $C[m][n]$, where $C[i][v]$ represents the count of MIS neighbors of $v$ in $\hat{G}_i$. 
We also maintain $\hat{\cal M}^+_i$ and $\hat{\cal M}^-_i$ storing the list of vertices inserted and deleted from $\hat{\cal M}_{i-1}$ to get $\hat{\cal M}_i$. After the first incorrect prediction $\hat{e}_{i^*+1}\neq e_{i^*+1}$, we maintain a list $E_m$ having edges with both end points in ${\cal M}_{i^*}$. Thereafter, $C[i^*][v]$ represents the count of neighbours of $v$ in ${\cal M}_{i^*}$, possibly violating $C[i^*][v]==0$ iff $v\in {\cal M}_{i^*}$ due to $E_m$. We additionally define $C_{m}[n]$ corresponding to each vertex as count offset, where the effective count becomes $C[i^*][v]+C_{m}[v]$ during the update. We also maintain $\cal M^-$ and $\cal M^+$ as vertices removed and added to ${\cal M}$. We also explicitly maintain $\cal M$ as the  MIS of the current graph.

\begin{algorithm}
\caption{$Insert(u,v)$: Update MIS after insertion of edge $(u,v)$ post prediction error.}
\label{alg:insert}
\begin{multicols}{2}
Add $(u,v)$ to $G$\;
Restore $\cal M$ to ${\cal M}_{i^*}$\;
\lIf{$u\in \cal M$}{$C[i^*][v]++$} 
\lIf{$v\in {\cal M}$}{$C[i^*][u]++$}
\BlankLine
\If{$u\in \cal M$ and $v\in \cal M$}{
Add $(u,v)$ to $E_m$\;
${\cal M}^-\gets V(E_m)\setminus $ MIS of $G(V(E_m),E_m)$\;
}
\BlankLine
$\cal M^+\gets \emptyset$\;

\ForAll{$x \in \cal M^-$}{
    \ForAll{$(x,y)\in E$}{
    $C_{m}[y]\gets C_{m}[y]-1$\;
    \lIf{$y\in V(E_m)$}{\textbf{continue}}
        \If{$C[i^*][y]+C_{m}[y] == 0$}{
            Add $y$ to $\cal M^+$\;
            \lForAll{$(y,z) \in E$}{
                $C_{m}[z]++$
            }
        }
    } 
}
${\cal M}\gets \cal M\cup M^+\setminus\cal M^-$\;
Restore values of $C_m[\cdot]$\;

\end{multicols}
\end{algorithm}

\subsection{Algorithm}
    We avoid repeatedly removing and inserting the same vertex from the MIS, by using the MIS of the subgraph induced by the harder set of edge insertions $E_m$. \\
    
    \noindent
    \textbf{Preprocessing.} We use the incremental MIS algorithm (\Cref{thm:dynMIS}) to process the predicted update sequence $\hat{U}$, maintaining the count of MIS neighbours for each vertex. After each update $\hat{e}_i$, we copy the counts of each $v\in V$ to $C[i][v]$. Again, the changes from $\hat{\cal M}_{i-1}$ to $\hat{\cal M}_{i}$ are updated in $\hat{\cal M}^+_i$ and $\hat{\cal M}^-_i$. We also initialize $E_m=\emptyset$ and $C_m[v]=0,v\in V$.\\

\noindent
\textbf{Update.}
While the update $e_i=\hat{e}_i$, we only update $\cal M$ using $\hat{\cal M}^+_i$ and $\hat{\cal M}^-_i$.
After the last such edge $e_{i^*}=\hat{e}_{i^*}$, where $e_{i^*+1}\neq \hat{e}_{i^*+1}$, we continue to use the last correctly predicted count $C[i^*][\cdot]$ for future processing. Now, on insertion of an edge $(u,v)$ (see \Cref{alg:insert}) we differentiate between \textit{easy} insertion ($u$ or $v\notin{\cal M}_{i^*}$) and \textit{hard} insertion (both $u,v\in {\cal M}_{i^*}$). The impact of every edge is updated directly on $C[i^*][\cdot]$, while the hard edges are collected in $E_m$. The hard edges violate the independence of $\cal M$, hence the MIS of $G(V(E_m),E_m)$ is computed, and the remaining vertices of $V(E_m)$ are added to $\cal M^-$. Since the removal of such vertices may potentially add some vertices to $\cal M$, we maintain the offset count $C_{m}[\cdot]$ to compute the same without affecting $C[i^*][\cdot]$.  
For each vertex $x\in \cal M^-$, the count of its neighbours is reduced. Any such neighbour $y$ whose effective count $C[i^*][y]+C_{m}[y]$ becomes zero is added to $\cal M^+$, and has counts of its neighbours incremented. Except if $y\in V(E_m)$, whose impact on the count of its neighbours is already accounted in $C[i^*][\cdot]$. Finally, $\cal M$ is updated using $\cal M^+$ and $\cal M^-$, and $C_m[\cdot]$ is restored so it can be used for the next update.

\begin{figure}[h!]
\centering


\subcaptionbox{Initial state}[0.30\linewidth]{%
\begin{tikzpicture}[scale=0.7, transform shape,
    every node/.style={circle, draw, minimum size=6mm, inner sep=0pt},
    dark/.style={fill=cyan!40, text=black},
    light/.style={fill=gray!20, text=black},
    thick
]

\node[dark,fill=cyan!40, label=above:{\scriptsize $1$}] (a)  at (0,3) {a};
\node[light, fill=cyan!40, label=above:{\scriptsize $1$}] (b) at (1.3,3.5) {b};
\node[light, label=above:{\scriptsize $1$}] (i) at (2.7,3) {i};
\node[light, label=above:{\scriptsize $1$}] (j) at (3.7,3.5) {j};
\node[light, fill=cyan!40, label=above:{\scriptsize $0$}] (k) at (2.7, 4.7) {k};
\node[dark,fill=cyan!40, label=above:{\scriptsize $2$}] (c)  at (0.5,1.5) {c};
\node[light, fill=cyan!40, label=above:{\scriptsize $2$}] (d) at (2,1) {d};
\node[light, fill=cyan!40, label=above:{\scriptsize $2$}] (e) at (0,0.5) {e};
\node[light, label=above:{\scriptsize $3$}] (l) at (3.2,0.9) {l};
\node[dark, fill=cyan!40, label=above:{\scriptsize $0$}] (m) at (3.5,2) {m};

\node[dark, fill=cyan!40, label=above:{\scriptsize $1$}] (g)  at (0, -1) {g};
\node[light, fill=cyan!40, label=above:{\scriptsize $2$}] (h) at (1.3, -1.5) {h};
\node[dark, fill=cyan!40, label=above:{\scriptsize $1$}] (f)  at (1.4, -0.2) {f};
\node[light, label=above:{\scriptsize $2$}] (n) at (3, -0.5) {n};

\draw (c) -- (d) [dashed];
\draw (d) -- (e) [dashed];
\draw (e) -- (c) [dashed];
\draw (d) -- (l);
\draw (l) -- (m);

\draw (a) -- (b) [dashed];
\draw (b) -- (l);
\draw (d) -- (n);
\draw (b) -- (i);
\draw (i) -- (j);
\draw (j) -- (k);
\draw (i) -- (l);

\draw (g) -- (h) [dashed];
\draw (f) to[bend left=15] (h) [dashed];
\draw (h) -- (n);

\draw (0.9,1) ellipse (1.6cm and 4.4cm);

\end{tikzpicture}
}%
\subcaptionbox{MIS of $G(V(E_m),E_m)$}[0.30\linewidth]{%
\begin{tikzpicture}[scale=0.7, transform shape,
    every node/.style={circle, draw, minimum size=6mm, inner sep=0pt},
    dark/.style={fill=cyan!40, text=black},
    light/.style={fill=gray!20, text=black},
    thick
]

\node[dark,fill=cyan!40, label=above:{\scriptsize $1$}] (a)  at (0,3) {a};
\node[light,label=above:{\scriptsize $1$}] (b) at (1.3,3.5) {b};
\node[light,label=above:{\scriptsize $1$}] (i) at (2.7,3) {i};
\node[light, label=above:{\scriptsize $1$}] (j) at (3.7,3.5) {j};
\node[light, fill=cyan!40, label=above:{\scriptsize $0$}] (k) at (2.7, 4.7) {k};
\node[light,label=above:{\scriptsize $2$}] (c)  at (0.5,1.5) {c};
\node[light,label=above:{\scriptsize $2$}] (d) at (2,1) {d};
\node[light, fill=cyan!40, label=above:{\scriptsize $2$}] (e) at (0,0.5) {e};
\node[light, label=above:{\scriptsize $3$}] (l) at (3.2,0.9) {l};
\node[dark, fill=cyan!40, label=above:{\scriptsize $0$}] (m) at (3.5,2) {m};

\node[light,label=above:{\scriptsize $1$}] (g)  at (0, -1) {g};
\node[light, fill=cyan!40, label=above:{\scriptsize $2$}] (h) at (1.3, -1.5) {h};
\node[light,label=above:{\scriptsize $1$}] (f)  at (1.4, -0.2) {f};
\node[light, label=above:{\scriptsize $2$}] (n) at (3, -0.5) {n};

\draw (c) -- (d);
\draw (d) -- (e);
\draw (e) -- (c);
\draw (d) -- (l);
\draw (l) -- (m);
\draw (b) -- (l);
\draw (d) -- (n);
\draw (a) -- (b);
\draw (b) -- (i);
\draw (i) -- (j);
\draw (j) -- (k);
\draw (i) -- (l);

\draw (g) -- (h);
\draw (f) to[bend left=15] (h);
\draw (h) -- (n);

\draw (0.9,1) ellipse (1.6cm and 4.4cm);

\end{tikzpicture}
}
\subcaptionbox{Processing $\cal M^-$}[0.30\linewidth]{%
\begin{tikzpicture}[scale=0.7, transform shape,
    every node/.style={circle, draw, minimum size=6mm, inner sep=0pt},
    dark/.style={fill=cyan!40, text=black},
    light/.style={fill=gray!20, text=black},
    thick
]

\node[dark,fill=cyan!40, label=above:{\scriptsize $0$}] (a)  at (0,3) {a};
\node[light,label=above:{\scriptsize $1$}] (b) at (1.3,3.5) {b};
\node[light, fill=cyan!40, label=above:{\scriptsize $0$}] (i) at (2.7,3) {i};
\node[light, label=above:{\scriptsize $2$}] (j) at (3.7,3.5) {j};
\node[light, fill=cyan!40, label=above:{\scriptsize $0$}] (k) at (2.7, 4.7) {k};
\node[light, label=above:{\scriptsize $1$}] (c)  at (0.5,1.5) {c};
\node[light, label=above:{\scriptsize $1$}] (d) at (2,1) {d};
\node[light, fill=cyan!40, label=above:{\scriptsize $0$}] (e) at (0,0.5) {e};
\node[light, label=above:{\scriptsize $2$}] (l) at (3.2,0.9) {l};
\node[dark, fill=cyan!40, label=above:{\scriptsize $0$}] (m) at (3.5,2) {m};

\node[light, label=above:{\scriptsize $1$}] (g)  at (0, -1) {g};
\node[light, fill=cyan!40, label=above:{\scriptsize $0$}] (h) at (1.3, -1.5) {h};
\node[light, label=above:{\scriptsize $1$}] (f)  at (1.4, -0.2) {f};
\node[light, label=above:{\scriptsize $1$}] (n) at (3, -0.5) {n};

\draw (c) -- (d);
\draw (d) -- (e);
\draw (e) -- (c);
\draw (d) -- (l);
\draw (l) -- (m);
\draw (d) -- (n);
\draw (b) -- (l);
\draw (a) -- (b);
\draw (b) -- (i);
\draw (i) -- (j);
\draw (j) -- (k);
\draw (i) -- (l);

\draw (g) -- (h);
\draw (f) to[bend left=15] (h);
\draw (h) -- (n);

\draw (0.9,1) ellipse (1.6cm and 4.4cm);

\end{tikzpicture}
}

\caption{Example of processing $E_m$ (dashed), $G(V(E_m),E_m)$ (ellipse) showing effective count.}
\label{fig:types-edges-inc}
\end{figure}
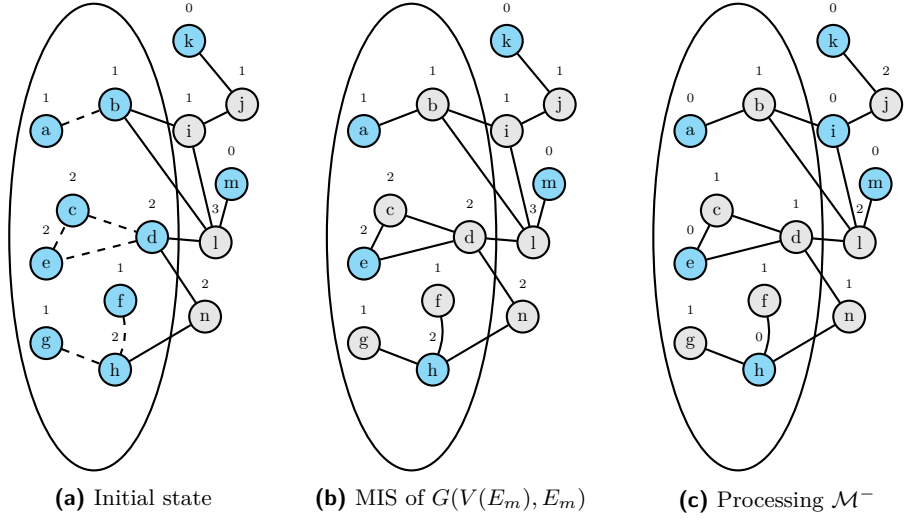

Consider \Cref{fig:types-edges-inc}~(a), we first apply the updates incrementing counts, with the hard updates $E_m$ shown as dashed. The MIS of the induced graph $G(V(E_m),E_m)$ is then computed as shown in \Cref{fig:types-edges-inc}~(b) to ensure independence. Thereafter, we reduce counts of neighbours of ${\cal M}^-=\{b,c,d,f,g\}$, resulting in an effective count zero for $i$, adding it to ${\cal M^+}$. Note that after processing $\cal M^-$ the effective count of all vertices in MIS of $G(V(E_m),E_m)$ becomes zero, and the remaining vertices of $V(E_m)$ remains non-zero.

\subsection{Correctness}
Again, the correctness follows from the invariant: vertices $v\in V$ with effective count $C[i^*][v]+C_m[v]==0$ iff $v\in \cal M$ before $C_m[\cdot]$ is restored. The independence of ${\cal M}_{i^*}$ is not violated by the easy updates, which simply update the count. It is violated due to $E_m$, which is restored by computing the MIS of the corresponding induced graph. 
However, removing $\cal M^-$ from $\cal M$ potentially violates the maximality, which is evaluated by updating the effective count of its neighbours. Adding the corresponding vertices with zero effective count to $\cal M^+$ restores the maximality. Note that reducing the counts of neighbours of $x\in\cal M^-$ also restores the effective count of vertices in the MIS of $G(V(E_m),E_m)$ to zero as expected, as they remain in $\cal M$. This count was earlier increased when the edges in $E_m$ were first added to the graph. Finally, note that a vertex in $\cal M^-$ cannot get its effective count zero, as the corresponding MIS neighbours in $G(V(E_m),E_m)$ are still in the MIS currently.

\subsection{Analysis}
The preprocessing again requires $O(mn)$ time and space to copy $C[i][v]$ for each vertex $v$ corresponding to every edge $\hat{e}_i$. Similarly, to compute $\hat{\cal M}^+_i,\hat{\cal M}^-_i$ we require $O(m)$ preprocessing time and space as the adjustment complexity is  amortized $O(1)$ per update (\Cref{thm:dynMIS}). While the prediction is true, each update is processed in amortized $O(1)$ time requiring total $O(m)$ update time again due to  $O(1)$ adjustment complexity. Thereafter, for each of the $\eta_e$ edges, the update time is dominated by $|E_m|$ for computing the MIS of $G(V(E_m),E_m)$, requiring $O(\eta_e)$ time. Further, it requires $\deg(v)$ for each $v\in \cal M^+\cup \cal M^-$ to update the count of its neighbours and later restore $C_m[\cdot]$, which is $O(\eta^*_v)$ by definition. 
Also, $\cal M$ can be restored in $O(\eta_v)$ as $\eta_v$ is the maximum of current and previous $\cal M^+\cup \cal M^-$.
This results in worst case update time $O(\eta_e+\eta^*_v)$ and total $O(m+\eta^2_e+\eta_e\eta^*_v)$ update time.

\incMIS*

\begin{remark} The above bound is for imperfect prediction only, while for perfect prediction we have optimal bound. The algorithm satisfies the desired properties of consistency, robustness and smooth degradation for the worst case update time since $\eta^*_v,\eta_e\leq 2m$. Note that any algorithm maintaining $\cal M$ explicitly always pays possibly $\omega(1)$ adjustment complexity of MIS, making our algorithm consistent for perfect predictions. However, it comes at the expense of significantly higher $O(mn)$ space and total time $O(m^2)$ for worst prediction, compared to the $O(m)$ space and $O(m\sqrt{m})$ total time of the incremental MIS algorithm (\Cref{thm:dynMIS}). 
\end{remark}

\section{Improved Incremental Algorithm}
In this section, we use the same compact data structure (\Cref{sec:impDec}) to reduce preprocessing time and space to $O(m\sqrt{m})$. We also avoid recomputing MIS of $G(V(E_m),E_m)$ from scratch during each update and use the incremental MIS algorithm instead, reducing the total update time to $O(m+\eta_e\sqrt{\eta_e}+\eta_e\eta_v^*+n\log \frac{m}{n})$.

\subsection{Data Structure}
We store the counts compactly in an array $C'[n]$ of dynamic arrays (similar to \Cref{sec:impDec}), adding $(i,c)$ to $C'[v]$ if count of $v$ changes on processing $\hat{e}_i$. Again, each $C'[v]$ is sorted corresponding to $i$. 
We maintain $\hat{\cal M}^+_i$ and $\hat{\cal M}^-_i$ storing the list of vertices inserted and deleted from $\hat{\cal M}_{i-1}$ to get $\hat{\cal M}_i$. 
After the first incorrect prediction $\hat{e}_{i^*+1}\neq e_{i^*+1}$, we maintain a list $E_m$ having edges with both end points in ${\cal M}_{i^*}$. We explicitly maintain $G_m=(V(E_m),E_m)$ incrementally under insertion of edges and vertices as $E_m$ is updated. 
Thereafter, $C'[v]$'s entry $(i,c_v)$ with last $i\leq i^*$ stores the count of neighbours in ${\cal M}_{i^*}$,  possibly violating $c_v==0$ iff $v\in {\cal M}_{i^*}$ due to $E_m$. The corresponding index in $C'[v]$ of the entry is stored in $j_v$.
Also, $C_{m}[n]$ corresponding to each vertex is the count offset, where the effective count becomes $c_v+C_{m}[v]$ during the update. We also maintain $\cal M^+$ and $\cal M^-$ as vertices added or removed from ${\cal M}$.  We also explicitly maintain $\cal M$ as the  MIS of the current graph.

\subsection{Algorithm}
We avoid repeatedly computing the MIS of $G(V(E_m),E_m)$ from scratch using the extended incremental MIS algorithm (\Cref{col:extInc}). Also, we use binary search wherever $C'[n]$ is required once for each vertex (similar to \Cref{alg:delete-2}). \\

    \noindent
    \textbf{Preprocessing.} 
We use the incremental MIS algorithm (\Cref{thm:dynMIS}) to process the predicted sequence $\hat{U}$, maintaining the count of MIS neighbours of each vertex. On processing $\hat{e}_i$, if the count of $v$ changes to $c$ we add  $(i,c)$ to $C'[v]$, implicitly sorting $C'[v]$. Again, the corresponding changes from $\hat{\cal M}_{i-1}$ to $\hat{\cal M}_{i}$ are updated in $\hat{\cal M}^+_i$ and $\hat{\cal M}^-_i$. We also initialize $E_m=\emptyset$, $C_m[v]=0,j_v=-1$ for each $v\in V$.\\

\noindent
\textbf{Update.}
While the update $e_i=\hat{e}_i$, we only update $\cal M$ using $\hat{\cal M}^+_i$ and $\hat{\cal M}^-_i$.
Further, after the last such edge $e_{i^*}=\hat{e}_{i^*}$ where $e_{i^*+1}\neq \hat{e}_{i^*+1}$, we continue to use the counts corresponding to $i^*$. This is computed using binary search once for each vertex $v$, by searching for the highest $(i,c_v)\in C'[v]$ where $i\leq i^*$. Now, on insertion of an edge $(u,v)$ (see \Cref{alg:insert-2}) we again differentiate between \textit{easy} insertion ($u$ or $v\notin{\cal M}_{i^*}$) and \textit{hard} insertion (both $u,v\in {\cal M}_{i^*}$). The impact of every edge is updated directly on their count in $C'[\cdot]$, while the hard edges are also collected in $E_m$. The hard edges violate the independence of $\cal M$, hence the MIS of $G(V(E_m),E_m)$ is maintained incrementally whenever $E_m$ changes. The remaining vertices of $V(E_m)$ are added to $\cal M^-$. Since the removal of such vertices may potentially add some vertices to $\cal M$, we maintain the offset count $C_{m}[\cdot]$ to compute the same without affecting $C'[\cdot]$.  
For each vertex $x\in \cal M^-$, the count of its neighbours is reduced. Any such neighbour $y$ whose effective count $c_y+C_{m}[y]$ becomes zero is added to $\cal M^+$, and has the count of its neighbours incremented. Note that in the absence of a valid $j_y$, it needs to be computed using binary search. Except if $y\in V(E_m)$, whose impact on the count of its neighbours is already accounted in $C'[\cdot]$. Finally, $\cal M$ is updated using $\cal M^+$ and $\cal M^-$, and $C_m[\cdot]$ is restored so it can be used for the next update.

\begin{algorithm}
\caption{$InsertI(u,v)$: Update MIS after insertion of $(u,v)$ post prediction error.}
\label{alg:insert-2}
\begin{multicols}{2}
Add $(u,v)$ to $G$\;
Restore $\cal M$ to ${\cal M}_{i^*}$\;
\If{$j_u==-1$}{
$j_u\gets$ last $(i,\cdot)\in C'[u]$ with $i\leq i^*$\;}
\If{$j_v==-1$}{
$j_v\gets$ last $(i,.)\in C'[v]$ with $i\leq i^*$\;}
$(i_u,c_u)\gets C'[u][j_u]$\;
$(i_v,c_v)\gets C'[v][j_v]$\;
\BlankLine

\lIf{$u\in {\cal M}$}{ $c_v++$, update $C'[v][j_v]$}
\lIf{$v\in \cal M$}{$c_u++$, update $C'[u][j_u]$}

\If{$u\in \cal M$ and $v\in \cal M$}
{
Add $(u,v)$ to $E_m$\; 
${\cal M}^-\gets V(E_m)\setminus$ IncMIS($G_m,(u,v)$)\; 
}

$\cal M^+\gets \emptyset$\;

\ForAll{$x \in \cal M^-$}{
    \ForAll{$(x,y)\in E$}{
    $C_{m}[y]\gets C_{m}[y]-1$\;
    \lIf{$y\in V(E_m)$}{\textbf{continue}}
        \If{$j_y==-1$}{
$j_y\gets$ last $(i,\cdot)\in C'[y]$ with $i\leq i^*$\;}
        $(i_y,c_y)\gets C'[y][j_y]$\;
        \If{$c_y+C_{m}[y] == 0$}{
            Add $y$ to $\cal M^{+}$\;
            \lForAll{$(y,z) \in E$}{
                $C_{m}[z]++$
            }
        }
    } 
}
${\cal M}\gets \cal M\cup M^+\setminus\cal M^-$\;

Restore values of $C_m[\cdot]$\;

\end{multicols}
\end{algorithm}

\subsection{Correctness}
The correctness follows from that of \Cref{alg:insert}, since we are only using alternate data structures to store count and maintain the MIS of $G(V(E_m),E_m)$ efficiently. 

\subsection{Analysis}

We first analyze the size of $C'[n]$. The incremental MIS algorithm (\Cref{thm:dynMIS}) explicitly maintains the count of each vertex. Hence, the number of changes to all the counts (and hence total size of $C'[\cdot]$) is bounded by its total time $O(m\sqrt{m})$.
Now, the preprocessing also requires $O(m\sqrt{m})$ time due to the incremental MIS algorithm, thereby computing $\hat{\cal M}^+_i,\hat{\cal M}^-_i$ and $C'[n]$. Again, while the prediction is true, each update requires amortized $O(1)$ time (due to amortized $O(1)$ adjustment complexity), resulting in total $O(m)$ time. 
Thereafter, for each of the $\eta_e$ edges, the update requires incrementally maintaining MIS of $G(V(E_m),E_m)$ taking worst case $O(|E_m|)=O(\eta_e)$ and amortized $O(\sqrt{|E_m|})=O(\sqrt{\eta_e})$ time (\Cref{col:extInc}).
Further, we require $\deg(v)$ for each $v\in \cal M^+\cup \cal M^-$ to update the count of its neighbours and later restore $C_m[\cdot]$, which is $O(\eta^*_v)$ by definition. Again, $\cal M$ can be restored in $O(\eta_v)$ as $\eta_v$ is the maximum of current and previous $\cal M^+\cup \cal M^-$.
Also, the binary search is required only once for each vertex $v$, requiring $O(\log \deg(v))$ time, taking total $O(n\log \frac{m}{n})$ (see \Cref{sec:impDec}). During an update, in the worst case it may be required for each neighbour of $x\in\cal M^-$ and $u,v$, requiring total $O(\log n+\mathbf{n}_v\log n)$ time.
 Thus, we have worst case $O(\log n + \eta_e+\eta^*_v+\mathbf{n}_v\log n)$ and total $O(m+\eta_e\sqrt{\eta_e} +\eta_e\eta^*_v+n\log\frac{m}{n})$ update times.

\incMISimp*

\begin{remark}
The improved algorithm requires less space while worst case bounds are still robust. However, the total time is robust except for the $O(\eta_e\eta^*_v)$ factor, which can be $O(m^2)$ for the worst predictions.   
\end{remark}

\section{Fully Dynamic MIS with Prediction}
We demonstrate how predicted updates are leveraged to improve the worst case update time of fully dynamic MIS from $O(m)$ to $O(\eta_e+\eta^*_v)$, using $O(un)$ preprocessing time and space.

\subsection{Data Structure}
We store a count array $C[u][n]$, where $C[i][v]$ represents the count of MIS neighbors of $v$ in $\hat{G}_i$.
We also maintain $\hat{\cal M}^+_i$ and $\hat{\cal M}^-_i$ storing the list of vertices inserted and deleted from $\hat{\cal M}_{i-1}$ to get $\hat{\cal M}_i$. After the first incorrect prediction $\langle\hat{e}_{i^*+1},\hat{t}_{i^*+1}\rangle\neq\langle e_{i^*+1},t_{i^*+1}\rangle$, we maintain three lists $E_m,E_i$ and $E_d$. The inserted edges with both end points in ${\cal M}_{i^*}$ are stored in $E_m$, and remaining inserted edges are stored in $E_i$. Also, the deleted edges are stored in $E_d$.  
We additionally define $C_{m}[n]$ corresponding to each vertex as count offset due to $E_i,E_m$ and $E_d$, where the effective count becomes $C[i^*][v]+C_{m}[v]$ during the update. We also maintain $\cal M^-$ and $\cal M^+$ as vertices removed and added to ${\cal M}$. We also explicitly maintain $\cal M$ as the  MIS of the current graph.

\begin{algorithm}
\caption{$update(\langle (u,v),t\rangle)$: Update MIS  for update $(t)$ of $(u,v)$ post prediction error.}
\label{alg:fully-dynamic}
\begin{multicols}{2}
\BlankLine
Restore ${\cal M}\gets {\cal M}_{i^*}$, $G\gets G_{i^*}$ \;
\uIf{$t==`-'$}{
\uIf{$(u,v)\in E_i\cup E_m$}{remove $(u,v)$ from $E_i$ and $E_m$\;}
\lElse{add $(u,v)$ to $E_d$}
}\Else{
\uIf{$(u,v)\in E_d$}{remove $(u,v)$ from $E_d$}
\Else{
\uIf{$u\in \cal M$ and $v\in \cal M$}{
add $(u,v)$ to $E_m$}
\lElse{add $(u,v)$ to $E_i$}
}
}
\BlankLine
\For{$(x,y)\in E_i\cup E_m$}{
add $(x,y)$ to $G$\;
\lIf{$x\in \cal M$}{$C_m[y]++$} 
\lIf{$y\in \cal M$}{$C_m[x]++$}
}
${\cal M}^-\gets V(E_m)\setminus $ MIS of $G(V(E_m),E_m)$\;
${\cal M}^+\gets \emptyset$\;

\ForAll{$x \in \cal M^-$}{
    \ForAll{$(x,y)\in E$}{
    $C_{m}[y]\gets C_{m}[y]-1$\;
    \lIf{$y\in V(E_m)$}{\textbf{continue}}
        \If{$C[i^*][y]+C_{m}[y] == 0$}{
            Add $y$ to $\cal M^{+}$\;
            \lForAll{$(y,z) \in E$}{
                $C_{m}[z]++$
            }
        }
    } 
}

\ForAll{$(x,y) \in E_d$}{
    remove $(x,y)$ from $G$\;
    \If{$C[i^*][y]+C_{m}[y] == 0$} 
        {Exchange $x$ and $y$\;}
    \lIf{$C[i^*][x]+C_{m}[x] \neq 0$}{
        {\textbf{continue}}
    }
    \tcc{$x\in $ MIS and $y\notin$ MIS}
    $C_{m}[y] \gets C_{m}[y]-1$\;
    \If{$C[i^*][y]+C_{m}[y] == 0$}{
    Add $y$ to $\cal M^+$\;
    \lForAll{$(y,z) \in E$}{
                $C_{m}[z]++$
    }  
    }
}
${\cal M}\gets \cal M\cup M^+\setminus\cal M^-$\;
Restore values of $C_m[\cdot]$\;

\end{multicols}
\end{algorithm}

\subsection{Algorithm} 
We again avoid removing and inserting the same vertex from the MIS, by first processing the insertion (similar to \Cref{alg:insert}) and then deletions. We also avoid the same edges being inserted and deleted by maintaining disjoint $E_m,E_i$ and $E_d$.
\\

\noindent
\textbf{Preprocessing.} 
 We use the fully dynamic MIS algorithm (\Cref{thm:dynMIS}) to process the predicted update sequence $\hat{U}$, maintaining the count of MIS neighbours for each vertex. After each update $\langle\hat{e}_i,\hat{t}_i\rangle$, we copy the counts of each $v\in V$ to $C[i][v]$. Again, the corresponding changes from $\hat{\cal M}_{i-1}$ to $\hat{\cal M}_{i}$ are updated in $\hat{\cal M}^+_i$ and $\hat{\cal M}^-_i$. We also initialize $E_m,E_i,E_d=\emptyset$ and $C_m[v]=0,v\in V$.\\

\noindent
\textbf{Update.} 
While the update $e_i=\hat{e}_i$, we only update $\cal M$ using $\hat{\cal M}^+_i$ and $\hat{\cal M}^-_i$.
After the last such edge $\langle e_{i^*},t_{i^*}\rangle=\langle \hat{e}_{i^*},\hat{t}_{i^*}\rangle$, where $\langle e_{i^*+1},t_{i^*+1}\rangle\neq \langle \hat{e}_{i^*+1},\hat{t}_{i^*+1}\rangle$, we continue to use the last correctly predicted count $C[i^*][\cdot]$ for future processing. Now, for an update $\langle(u,v),t\rangle$ (see \Cref{alg:fully-dynamic}) we maintain disjoint set of edges namely $E_m$ for \textit{hard} insertions (both $u,v\in {\cal M}_{i^*}$), $E_i$ for remaining easy insertions, and $E_d$ for deletions. We update the impact of edges insertions $E_i,E_m$ on the offset count $C_m[\cdot]$, without affecting $C[i^*][\cdot]$. The hard edge insertions $E_m$ violate the independence of $\cal M$, hence the MIS of $G(V(E_m),E_m)$ is computed, and the remaining vertices of $V(E_m)$ are added to $\cal M^-$. Since the removal of such vertices may potentially add some vertices to $\cal M$, we update the offset count $C_{m}[\cdot]$ to compute the same.  
For each vertex $x\in \cal M^-$, the count of its neighbours is reduced. Any such neighbour $y$ whose effective count $C[i^*][y]+C_{m}[y]$ becomes zero is added to $\cal M^+$, and has counts of its neighbours incremented. Except if $y\in V(E_m)$, whose impact on the count of its neighbours is already accounted in $C[i^*][\cdot]$. 
Now, the impact of $E_i$ and $E_m$ is completely accounted for in the effective count and the current MIS $\cal M\cup\cal M^+\setminus \cal M^-$. We then process $E_d$ (similar to decremental MIS algorithm in \Cref{thm:dynMIS}), updating counts on deleted edges if incident on the current MIS. Again, if effective count becomes zero the vertex is added to $\cal M^+$ and the count of its neighbours are increased.
Finally, $\cal M$ is updated using $\cal M^+$ and $\cal M^-$, and $C_m[\cdot]$ is restored so it can be used for the next update. 

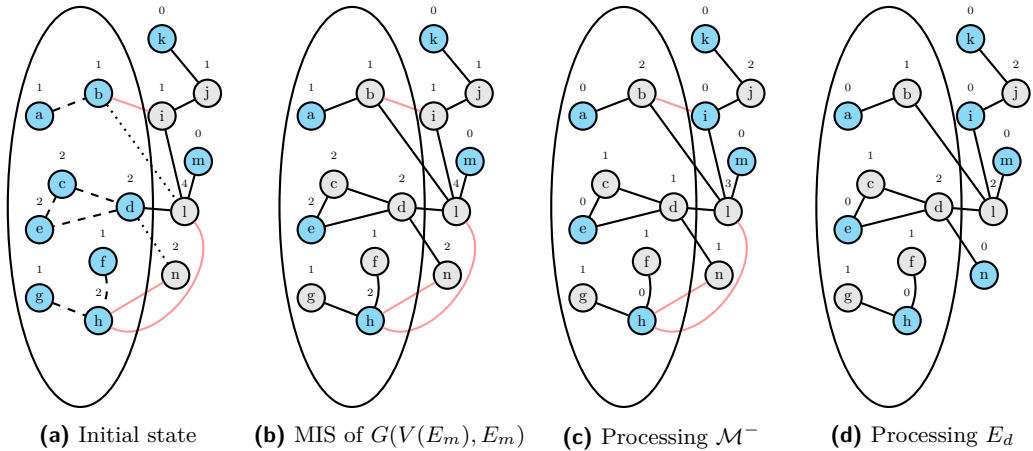
\begin{figure}[!ht]
\centering

\begin{minipage}[t]{0.24\textwidth}
\centering

\begin{tikzpicture}[scale=0.6, transform shape,
     every node/.style={circle, draw, minimum size=6mm, inner sep=0pt},
    dark/.style={fill=cyan!40, text=black},
    light/.style={fill=gray!20, text=black},
    thick
]
\node[dark, label=above:{\scriptsize $1$}] (a)  at (0,3) {a};
\node[dark, label=above:{\scriptsize $1$}] (b) at (1.3,3.5) {b};
\node[light, label=above:{\scriptsize $1$}] (i) at (2.7,3) {i};
\node[light, label=above:{\scriptsize $1$}] (j) at (3.7,3.5) {j};
\node[dark, label=above:{\scriptsize $0$}] (k) at (2.7, 4.7) {k};
\node[dark, label=above:{\scriptsize $2$}] (c)  at (0.5,1.5) {c};
\node[dark, label=above:{\scriptsize $2$}] (d) at (2,1) {d};
\node[dark, label=above:{\scriptsize $2$}] (e) at (0,0.5) {e};
\node[light, label=above:{\scriptsize $4$}] (l) at (3.2,0.9) {l};
\node[dark, label=above:{\scriptsize $0$}] (m) at (3.5,2) {m};

\node[dark, label=above:{\scriptsize $1$}] (g)  at (0, -1) {g};
\node[dark, label=above:{\scriptsize $2$}] (h) at (1.3, -1.5) {h};
\node[dark, label=above:{\scriptsize $1$}] (f)  at (1.4, -0.2) {f};
\node[light, label=above:{\scriptsize $2$}] (n) at (3, -0.5) {n};

\draw (c) -- (d) [dashed];
\draw (d) -- (e) [dashed];
\draw (e) -- (c) [dashed];
\draw (d) -- (l);
\draw (l) -- (m);

\draw (a) -- (b) [dashed];
\draw[dotted] (b) -- (l);
\draw[dotted](d) -- (n);
\draw[red!40] (b) -- (i) ;
\draw (i) -- (j);
\draw (j) -- (k);
\draw (i) -- (l);
\draw[red!40] (l) to[bend left=80] (h);

\draw (g) -- (h) [dashed];
\draw (f) to[bend left=15] (h) [dashed];
\draw[red!40] (h) -- (n);

\draw (0.9,1) ellipse (1.6cm and 4.4cm);
\end{tikzpicture}
\subcaption{Initial state}
\end{minipage}\hfill%
\begin{minipage}[t]{0.26\textwidth}
\centering
\begin{tikzpicture}[scale=0.6, transform shape,
    every node/.style={circle, draw, minimum size=6mm, inner sep=0pt},
    dark/.style={fill=cyan!40, text=black},
    light/.style={fill=gray!20, text=black},
    thick
]
\node[dark,fill=cyan!40, label=above:{\scriptsize $1$}] (a)  at (0,3) {a};
\node[light,label=above:{\scriptsize $1$}] (b) at (1.3,3.5) {b};
\node[light, label=above:{\scriptsize $1$}] (i) at (2.7,3) {i};
\node[light, label=above:{\scriptsize $1$}] (j) at (3.7,3.5) {j};
\node[light, fill=cyan!40, label=above:{\scriptsize $0$}] (k) at (2.7, 4.7) {k};
\node[light,label=above:{\scriptsize $2$}] (c)  at (0.5,1.5) {c};
\node[light,label=above:{\scriptsize $2$}] (d) at (2,1) {d};
\node[light, fill=cyan!40, label=above:{\scriptsize $2$}] (e) at (0,0.5) {e};
\node[light, label=above:{\scriptsize $4$}] (l) at (3.2,0.9) {l};
\node[dark, fill=cyan!40, label=above:{\scriptsize $0$}] (m) at (3.5,2) {m};

\node[light,label=above:{\scriptsize $1$}] (g)  at (0, -1) {g};
\node[light, fill=cyan!40, label=above:{\scriptsize $2$}] (h) at (1.3, -1.5) {h};
\node[light,label=above:{\scriptsize $1$}] (f)  at (1.4, -0.2) {f};
\node[light, label=above:{\scriptsize $2$}] (n) at (3, -0.5) {n};

\draw (c) -- (d);
\draw (d) -- (e);
\draw (e) -- (c);
\draw (d) -- (l);
\draw[red!40] (b) -- (i);
\draw (l) -- (m);
\draw (b) -- (l);
\draw (d) -- (n);
\draw (a) -- (b);
\draw[red!40] (l) to[bend left=80] (h);
\draw (i) -- (j);
\draw (j) -- (k);
\draw (i) -- (l);

\draw (g) -- (h);
\draw (f) to[bend left=15] (h);
\draw[red!40] (h) -- (n);

\draw (0.9,1) ellipse (1.6cm and 4.4cm);
\end{tikzpicture}
\subcaption{MIS of $G(V(E_m),E_m)$}
\end{minipage}\hfill%
\begin{minipage}[t]{0.24\textwidth}
\centering

\begin{tikzpicture}[scale=0.6, transform shape,
    every node/.style={circle, draw, minimum size=6mm, inner sep=0pt},
    dark/.style={fill=cyan!40, text=black},
    light/.style={fill=gray!20, text=black},
    thick
]
\node[dark,fill=cyan!40, label=above:{\scriptsize $0$}] (a)  at (0,3) {a};
\node[light,label=above:{\scriptsize $2$}] (b) at (1.3,3.5) {b};
\node[light, fill=cyan!40, label=above:{\scriptsize $0$}] (i) at (2.7,3) {i};
\node[light, label=above:{\scriptsize $2$}] (j) at (3.7,3.5) {j};
\node[light, fill=cyan!40, label=above:{\scriptsize $0$}] (k) at (2.7, 4.7) {k};
\node[light, label=above:{\scriptsize $1$}] (c)  at (0.5,1.5) {c};
\node[light, label=above:{\scriptsize $1$}] (d) at (2,1) {d};
\node[light, fill=cyan!40, label=above:{\scriptsize $0$}] (e) at (0,0.5) {e};
\node[light, label=above:{\scriptsize $3$}] (l) at (3.2,0.9) {l};
\node[dark, fill=cyan!40, label=above:{\scriptsize $0$}] (m) at (3.5,2) {m};

\node[light, label=above:{\scriptsize $1$}] (g)  at (0, -1) {g};
\node[light, fill=cyan!40, label=above:{\scriptsize $0$}] (h) at (1.3, -1.5) {h};
\node[light, label=above:{\scriptsize $1$}] (f)  at (1.4, -0.2) {f};
\node[light, label=above:{\scriptsize $1$}] (n) at (3, -0.5) {n};

\draw (c) -- (d);
\draw (d) -- (e);
\draw (e) -- (c);
\draw (d) -- (l);
\draw (l) -- (m);
\draw[red!40] (b) -- (i);
\draw (d) -- (n);
\draw (b) -- (l);
\draw (a) -- (b);
\draw[red!40] (l) to[bend left=80] (h);
\draw (i) -- (j);
\draw (j) -- (k);
\draw (i) -- (l);

\draw (g) -- (h);
\draw (f) to[bend left=15] (h);
\draw[red!40] (h) -- (n);

\draw (0.9,1) ellipse (1.6cm and 4.4cm);

\end{tikzpicture}
\subcaption{Processing $\mathcal M^-$}
\end{minipage}\hfill%
\begin{minipage}[t]{0.24\textwidth}
\centering

\begin{tikzpicture}[scale=0.6, transform shape,
    every node/.style={circle, draw, minimum size=6mm, inner sep=0pt},
    dark/.style={fill=cyan!40, text=black},
    light/.style={fill=gray!20, text=black},
    thick
]
\node[dark,fill=cyan!40, label=above:{\scriptsize $0$}] (a)  at (0,3) {a};
\node[light,label=above:{\scriptsize $1$}] (b) at (1.3,3.5) {b};
\node[light, fill=cyan!40, label=above:{\scriptsize $0$}] (i) at (2.7,3) {i};
\node[light, label=above:{\scriptsize $2$}] (j) at (3.7,3.5) {j};
\node[light, fill=cyan!40, label=above:{\scriptsize $0$}] (k) at (2.7, 4.7) {k};
\node[light, label=above:{\scriptsize $1$}] (c)  at (0.5,1.5) {c};
\node[light, label=above:{\scriptsize $2$}] (d) at (2,1) {d};
\node[light, fill=cyan!40, label=above:{\scriptsize $0$}] (e) at (0,0.5) {e};
\node[light, label=above:{\scriptsize $2$}] (l) at (3.2,0.9) {l};
\node[dark, fill=cyan!40, label=above:{\scriptsize $0$}] (m) at (3.5,2) {m};

\node[light, label=above:{\scriptsize $1$}] (g)  at (0, -1) {g};
\node[light, fill=cyan!40, label=above:{\scriptsize $0$}] (h) at (1.3, -1.5) {h};
\node[light, label=above:{\scriptsize $1$}] (f)  at (1.4, -0.2) {f};
\node[light, fill=cyan!40, label=above:{\scriptsize $0$}] (n) at (3, -0.5) {n};

\draw (c) -- (d);
\draw (d) -- (e);
\draw (e) -- (c);
\draw (d) -- (l);
\draw (l) -- (m);
\draw (d) -- (n);
\draw (b) -- (l);
\draw (a) -- (b);
\draw (i) -- (j);
\draw (j) -- (k);
\draw (i) -- (l);

\draw (g) -- (h);
\draw (f) to[bend left=15] (h);

\draw (0.9,1) ellipse (1.6cm and 4.4cm);
\end{tikzpicture}
\subcaption{Processing $E_d$}
\end{minipage}

\caption{Example of processing $E_i$ (dotted), $E_m$ (dashed) and $E_d$ (red) showing effective count.}
\label{fig:types-edges}
\end{figure}

Consider \Cref{fig:types-edges}~(a), the initial state after applying the updates $E_i$ (dotted) and $E_m$ (dashed) on the effective counts. The MIS of the induced graph $G(V(E_m),E_m)$ is then computed as shown in \Cref{fig:types-edges}~(b) to ensure independence. Thereafter, we reduce counts of neighbours of ${\cal M}^-=\{b,c,d,f,g\}$, resulting in zero effective count for $i$, adding it to ${\cal M^+}$, updating counts of its neighbours as shown in \Cref{fig:types-edges}~(c). Note that the processing of $E_i$ and $E_m$ is complete, updating the effective counts to zero for the MIS of $G(V(E_m),E_m)$.
Finally, after deleting  $E_d$ (red) counts are reduced accordingly, resulting in zero effective count for $n$, adding it to $\cal M^+$,  updating counts of its neighbours as shown in \Cref{fig:types-edges}~(d).

\subsection{Correctness}
Again, the correctness follows from that of \Cref{alg:insert} since we use the same algorithm followed by deletion using simple decremental algorithm (\Cref{thm:dynMIS}). The only concern is whether a vertex removed $\cal M^-$ can potentially be inserted back in $\cal M^+$ while processing $E_d$. However, again the MIS neighbours in $G(V(E_m),E_m)$ for such a vertex will continue to impact its effective count since $E_d$ and $E_m$ are disjoint.

\subsection{Analysis}
The preprocessing requires $O(un)$ time and space to copy $C[i][v]$ for each vertex $v$ corresponding to every edge $\hat{e}_i$ (\Cref{thm:dynMIS}). Similarly, to compute $\hat{\cal M}^+_i,\hat{\cal M}^-_i$ we require $O(u)$ preprocessing time and space as the adjustment complexity is amortized $O(1)$ per update (\Cref{thm:dynMIS}). While the prediction is true, each update is processed in amortized $O(1)$ time, requiring total $O(u)$ update time again due to  $O(1)$ adjustment complexity. Thereafter, for each of the $\eta'_e$ edges, the update time processes $E_m,E_i,E_d$ along with computing the MIS of $G(V(E_m),E_m)$, in $O(\eta_e)$ time. Further, it requires $\deg(v)$ for each $v\in \cal M^+\cup \cal M^-$ to update the count of its neighbours and later restore $C_m[\cdot]$, which is $O(\eta^*_v)$ by definition. This is because no vertex is both removed and inserted, having disjoint $\cal M^+,\cal M^-$ (as described previously).
Again, $\cal M$ can be restored in $O(\eta_v)$ as earlier, and $G$ can be restored in $O(\eta_e)$.
This results in worst case update time $O(\eta_e+\eta^*_v)$ and total $O(u+\eta'_e\eta_e+\eta'_e\eta^*_v)$ update time.

\fdMIS*

\begin{remark} The above bound is for imperfect prediction only, while for perfect prediction we have optimal bound. The algorithm satisfies the desired properties of consistency, robustness and smooth degradation for the worst case update time since $\eta^*_v,\eta_e\leq 2m$. Again, any algorithm maintaining $\cal M$ explicitly always pays possibly $\omega(1)$ adjustment complexity of MIS, making our algorithm consistent for perfect predictions. However, it comes at the expense of significantly higher $O(un)$ space and total time $O(um)$ for worst prediction, compared to the $O(m)$ space and $O(um^{2/3})$ total time of the fully dynamic MIS algorithm (\Cref{thm:dynMIS2}). 
\end{remark}

\begin{remark}
We cannot improve the preprocessing time and space using faster fully dynamic MIS algorithm with $O(m^{2/3})$ amortized update time. This is because we require the explicit $count$ being computed after each update in $C[u][n]$. 
\end{remark}

 \begin{remark}
    The processing of $E_d$ removes the benefit of updating $E_m,E_i,E_d$ directly into $C[i^*][\cdot]$ as in \Cref{sec:incMIS}, or using faster fully dynamic MIS for maintaining MIS of $G(V(E_m),E_m)$. This is because the vertices entering $\cal M^+$ while processing $E_d$, may be much smaller than even those edges of $E_d$ which makes some count zero. Hence, the total update time of $O(\eta'_e\eta_e)$ is inevitable with the current approach.
 \end{remark}

\section{Conclusion}
\label{sec:Conclusion}
We presented the first algorithms for maintaining a Maximal Independent Set in the prediction model, under incremental, decremental, and fully dynamic edge updates. Our algorithms exploit a predicted update sequence, which is preprocessed offline to accelerate online repair when real updates arrive. The resulting running times scale with prediction quality rather than with worst-case graph parameters, maintaining the desired properties of learning augmented algorithms: consistency, robustness, and smooth degradation.

However, this comes at the expense of higher space and total update time. We thus present space-efficient variants that reduce the space to $O(m)$ for decremental algorithm and  $O(m\sqrt{m})$ for incremental algorithm, at the expense of a logarithmic factor in the worst case update time. These algorithms also significantly improve the total update time. 

Note that our algorithms follow robustness for only worst case update time, but not total update time (except decremental for dense graphs). Notably, our algorithms employ a relatively simpler approach, without exploiting the classical vertex decomposition (eg, light and heavy) of the state-of-the-art fully dynamic MIS algorithms~\cite{AssadiOSS18,GuptaK21,GuptaKS26}. In the future, it would be interesting to see the impact of such approaches that are seemingly intrinsic to the Maximal Independent Set problem.

\bibliography{paper}



\end{document}